\documentclass[11pt]{book}

\usepackage[dvips]{epsfig}
\usepackage{amssymb}
\usepackage{multirow}
\usepackage{amsbsy}
\usepackage{pstricks}
\usepackage{amsmath}
\usepackage{bm}
\usepackage{newtxtext,newtxmath}
\usepackage{hyperref}
\catcode`\@=11

 \def\@normalsize{\@setsize\normalsize{13pt}\xipt\@xipt
   \abovedisplayskip 11pt plus3pt minus6pt
   \belowdisplayskip \abovedisplayskip
   \abovedisplayshortskip \z@ plus3pt
   \belowdisplayshortskip 6.6pt plus3.5pt minus3pt}

 \def\small{\@setsize\small{12pt}\xipt\@xipt
   \abovedisplayskip 10pt plus2pt minus5pt
   \belowdisplayskip \abovedisplayskip
   \abovedisplayshortskip \z@ plus3pt
   \belowdisplayshortskip 6pt plus3pt minus3pt
   \def\@listi{\topsep 6pt plus 2pt minus 2pt
     \parsep 3pt plus 2pt minus 1pt
     \itemsep \parsep}}

 \def\footnotesize{\@setsize\footnotesize{10pt}\ixpt\@ixpt
   \abovedisplayskip 8pt plus 2pt minus 4pt
   \belowdisplayskip \abovedisplayskip
   \abovedisplayshortskip \z@ plus 1pt
   \belowdisplayshortskip 4pt plus 2pt minus 2pt
   \def\@listi{\topsep 4pt plus 2pt minus 2pt
      \parsep 2pt plus 1pt minus 1pt
      \itemsep \parsep}}

 \def\scriptsize{\@setsize\scriptsize{9.5pt}\viiipt\@viiipt}
 \def\tiny{\@setsize\tiny{7pt}\vipt\@vipt}
 \def\large{\@setsize\large{14pt}\xiipt\@xiipt}
 \def\Large{\@setsize\Large{18pt}\xivpt\@xivpt}
 \def\LARGE{\@setsize\LARGE{22pt}\xviipt\@xviipt}
 \def\huge{\@setsize\huge{25pt}\xxpt\@xxpt}
 \def\Huge{\@setsize\Huge{30pt}\xxvpt\@xxvpt}
\def\section{\@startsection {section}{1}{\z@}%
{-1.5\baselineskip plus-1pt minus-3pt}{1\baselineskip plus1pt minus2pt}%
{\centering\normalsize\bf}}
\def\subsection{\@startsection{subsection}{2}{\z@}%
{-1\baselineskip plus-1pt minus-2pt}{1\baselineskip plus1pt minus2pt}%
{\normalsize\sc\noindent}}
\def\subsubsection{\@startsection{subsubsection}{3}{\z@}%
{-1\baselineskip plus-1pt minus-2pt}{1sp}{\normalsize\it\noindent}}
\def\paragraph{\@startsection{paragraph}{4}{\z@}%
{1\baselineskip plus1pt minus2pt}{-1em}{\normalsize\it\noindent}}
\let\subparagraph=\paragraph

\def\tableofcontents{\@restonecolfalse\if@twocolumn\@restonecoltrue
\onecolumn\fi\OSIDcont\@starttoc{con}\if@restonecol\twocolumn\fi}

\def\l@section{\@dottedtocline{1}{0em}{.66em}}

\def\thebibliography#1{\section*{{Bibliography}\@mkboth
 {BIBLIOGRAPHY}{BIBLIOGRAPHY}}\footnotesize\rm\list
 {[\arabic{enumi}]}{\settowidth\labelwidth{[#1]}\leftmargin\labelwidth
 \advance\leftmargin\labelsep\usecounter{enumi}}
 \def\newblock{\hskip .11em plus .33em minus -.07em}
 \sloppy\clubpenalty4000\widowpenalty4000
 \sfcode`\.=1000\relax}

\def\ps@myheadings{\let\@mkboth\@gobbletwo
\def\@oddhead{\hfil{\footnotesize\rm\rightmark}\hfil}
\def\@evenhead{\hfil{\footnotesize\rm\leftmark}\hfil}
\def\@oddfoot{\hfil{\footnotesize\sf\artid-\thepage}\hfil}
\def\@evenfoot{\hfil{\footnotesize\sf\artid-\thepage}\hfil}
\def\sectionmark##1{}\def\subsectionmark##1{}}

\def\artid{0000001}
\newcounter{paPer}     %
\def\EndpagE{\expandafter\pageref{\the\value{paPer}OpSy}}

\def\ps@osiD{\let\@mkboth\@gobbletwo

  \def\@oddfoot{\hfil{\footnotesize\sf\artid-\thepage}\hfil}
  \def\@evenhead{}\let\@evenfoot\@oddfoot}

\def\cite{\@ifnextchar [{\@tempswatrue\@Rcitex}{\@tempswafalse\@Rcitex[]}}

\def\@Rcitex[#1]#2{\if@filesw\immediate\write\@auxout{\string\citation{#2}}\fi
  \def\@citea{}\@cite{\@for\@citeb:=#2\do
    {\@citea\def\@citea{,\penalty\@m\,}\@ifundefined
       {b@\the\value{paPer}R\@citeb}{{\bf ?}\@warning
       {Citation `\@citeb' on page \thepage \space undefined}}%
\hbox{\csname b@\the\value{paPer}R\@citeb\endcsname}}}{#1}}

\long\def\@caption#1[#2]#3{\par\addcontentsline{\csname
  ext@#1\endcsname}{#1}{\protect\numberline{\csname
  the#1\endcsname}{\ignorespaces #2}}\begingroup
    \@parboxrestore
    \small                                        %    \normalsize
    \@makecaption{\csname fnum@#1\endcsname}{\ignorespaces #3}\par
  \endgroup}

\newtoks\@stequation

\def\subequations{\refstepcounter{equation}%
\edef\@savedequation{\the\c@equation}%
\@stequation=\expandafter{\theequation}%   %only want \theequation
\edef\@savedtheequation{\the\@stequation}% %expanded once
\edef\oldtheequation{\theequation}%
\setcounter{equation}{0}%
\def\theequation{\oldtheequation\alph{equation}}}%

\def\endsubequations{%
\setcounter{equation}{\@savedequation}%
\@stequation=\expandafter{\@savedtheequation}%
\edef\theequation{\the\@stequation}\global\@ignoretrue}

\catcode`\@=12
\let\Rlabel=\label
\let\Rbibitem=\bibitem
\let\Rref=\ref
\let\Rpageref=\pageref
\def\label#1{\expandafter\Rlabel{\the\value{paPer}R#1}}
\def\bibitem#1{\expandafter\Rbibitem{\the\value{paPer}R#1}}
\def\ref#1{\expandafter\Rref{\the\value{paPer}R#1}}
\def\pageref#1{\expandafter\Rpageref{\the\value{paPer}R#1}}

\def\thesection{\arabic{section}.}

\def\YYMm{\rule{0ex}{4em}}
\newtoks\TITsi
\newtoks\TITsii

\def\title#1{\def\TITs{\LARGE{\raggedright\noindent\YYMm #1%
\vskip8pt\par}}}
\def\author#1{\autMM{#1}\def\LHD{#1}}
\def\and{{\rm\lowercase{and}}}

\def\autMM#1{\TITsii={\vskip10pt\par\normalsize\rm\noindent #1\par}%
\TITsi=\expandafter{\TITs}\edef\TITs{\the\TITsi\the\TITsii}}

\def\address#1{\TITsii={\vskip6pt\par\footnotesize\sl\noindent #1\par}%
\TITsi=\expandafter{\TITs}%
\edef\TITs{\the\TITsi\the\TITsii}}

\def\received#1{\TITsii={\vskip10pt\par\small\rm\noindent(Received: #1)\par}%
\TITsi=\expandafter{\TITs}\edef\TITs{\the\TITsi\the\TITsii}}

\def\headtitle#1{\def\RHD{#1}}
\def\headauthor#1{\def\LHD{#1}}
\def\listas#1#2{\addcontentsline{con}{section}{{\sc #1: }{\rm #2}}}

\def\abst{{\bf Abstract.}}
\def\abstract#1{\TITs
       \vskip15pt\par\noindent
       {\footnotesize{\abst~} #1\vskip3pt\par}
       \markright{\RHD}
       \markboth{\LHD}{\RHD}}

\def\startpaper{%
       \cleardoublepage
       \setcounter{section}{0}
       \stepcounter{paPer}
       \setcounter{equation}{0}
       \setcounter{footnote}{0}
       \setcounter{figure}{0}
       \setcounter{table}{0}
       \def\theequation{\arabic{equation}}
       \def\thefootnote{\arabic{footnote}}
       \setcounter{defn}{0}
       \setcounter{thm}{0}
       \setcounter{lem}{0}
       \setcounter{prop}{0}
       \setcounter{rem}{0}
       \thispagestyle{osiD}}

\def\OSIDcont{\cleardoublepage\thispagestyle{empty}
       \markright{}\markboth{}{}
       \normalsize\rm
       \hspace*{\fill}{\large\rm
         Contents of the Volume \Volume, Number \Number}\hspace*{\fill}
       \par\vspace{1.5em}
       \par\noindent}

\def\endpaper{\expandafter\label{\the\value{paPer}OpSy}}

\def\1{{\mathchoice{\rm 1\mskip-4mu l}{\rm 1\mskip-4mu l}%
{\rm 1\mskip-4.5mu l}{\rm 1\mskip-5mu l}}}
\def\varkappa{\mbox{\bBB\char 123}}
\def\longhookrightarrow{\lhook\joinrel\relbar\joinrel\rightarrow}

\def\longhookUp{\lower6pt\hbox{\rotatebox{90}{$\longhookrightarrow$}}}

\def\theequation{\thesection\arabic{equation}}

\def\Myskip{\setlength{\baselineskip}{13pt}}
\def\text#1{\quad\mbox{\rm  #1 }\quad}

\def\SMP{\def\thefootnote{*}%
\def\thefootnote{\arabic{footnote}}\setcounter{footnote}{0}}

\input xy
\xyoption{all}

\InputIfFileExists{psfig.sty}{\typeout{^^Jpsfig.sty inputed...ok}}{\typeout{^^JWarning: psfig.sty could not be found.^^J}}
\begin{document}

\startpaper

\newcommand{\Mn}{M_n(\mathbb{C})}
\newcommand{\Mk}{M_k(\mathbb{C})}
\newcommand{\id}{\mbox{id}}
\newcommand{\ot}{{\,\otimes\,}}
\newcommand{{\Cd}}{{\mathbb{C}^d}}
\newcommand{\sbsigma}{{\mbox{\scriptsize \boldmath $\sigma$}}}
\newcommand{\sbalpha}{{\mbox{\scriptsize \boldmath $\alpha$}}}
\newcommand{\sbbeta}{{\mbox{\scriptsize \boldmath $\beta$}}}
\newcommand{\bsigma}{{\mbox{\boldmath $\sigma$}}}
\newcommand{\balpha}{{\mbox{\boldmath $\alpha$}}}
\newcommand{\bbeta}{{\mbox{\boldmath $\beta$}}}
\newcommand{\bmu}{{\mbox{\boldmath $\mu$}}}
\newcommand{\bnu}{{\mbox{\boldmath $\nu$}}}
\newcommand{\ba}{{\mbox{\boldmath $a$}}}
\newcommand{\bb}{{\mbox{\boldmath $b$}}}
\newcommand{\sba}{{\mbox{\scriptsize \boldmath $a$}}}
\newcommand{\MD}{\mathfrak{D}}
\newcommand{\sbb}{{\mbox{\scriptsize \boldmath $b$}}}
\newcommand{\sbmu}{{\mbox{\scriptsize \boldmath $\mu$}}}
\newcommand{\sbnu}{{\mbox{\scriptsize \boldmath $\nu$}}}
\def\oper{{\mathchoice{\rm 1\mskip-4mu l}{\rm 1\mskip-4mu l}%
{\rm 1\mskip-4.5mu l}{\rm 1\mskip-5mu l}}}
\def\<{\langle}
\def\>{\rangle}
\def\theequation{\thesection\arabic{equation}}

\title{Dissipative Dynamics in Open Fermionic Chains}
\author{A. I. Karanikas$^1$ and G. E. Pavlou$^2$}
\address{$^1$ National and Kapodistrian University of Athens, Physics Department, Nuclear \& Particle Physics Section Panepistimiopolis, Ilissia 15771 Athens, Greece}
\address{$^2$ Institute of Applied and Computational Mathematics (IACM), Foundation for Research and Technology Hellas (FORTH), IACM/FORTH, GR-71110 Heraklion, Greece}
\headauthor{A. I. Karanikas and G. E. Pavlou}
\headtitle{Dissipative Dynamics in Open Fermionic Chains}
\received{March 10, 2023}
\listas{A. I. Karanikas and G. E. Pavlou}{Dissipative Dynamics in Open Fermionic Chains}

\abstract{By merging the Feynman-Vernon’s approach with the out-of-equilibrium Keldysh-Schwinger formalism, we construct the reduced generating functional through which all the time-dependent correlation functions of an open fermionic system can be directly derived by applying the appropriate functional derivatives. As a concrete example, we investigate the transverse Ising model, we derive the covariance matrix at the steady state of the system and we investigate its critical behavior. }

\Myskip

%\pacs{03.65.Ud, 03.67.-a}

\SMP

\section{Introduction}
\label{intro}
\setcounter{equation}{0}

The unavoidable influence of the environment on open quantum systems is of great importance for understanding their physical properties and for handling their practical applications. When a system is embedded in a controlled or uncontrolled environment, induced decoherence is one of the major issues for storing and processing quantum information. Among these systems, for reasons of both technical and theoretical origin, fermionic and spin chains with short-range interactions, play a central role as their dynamics reveal a quite interesting complexity even at their simple 1D versions. Isolated systems, the Hamiltonian of which is quadratic in fermionic or bosonic degrees of freedom, have been extensively studied as they are exactly solvable and the structure of their ground state offers the explanation basis for highly non-trivial phenomena as quantum phase transitions of topological character \cite{Osborne2002,Vidal2003}. However, the embedding of such a system into a bath induces major changes in the behavior of the system’s correlations in a way that is not easy to calculate even if the isolated system is completely solvable. When a system is isolated, the structure of its ground state as well as the energy gap between this state and the excited ones has a central role for the system’s properties \cite{Sachdev2011}. In the case of an open system, the role of the ground state is played by the so-called steady state that is, the state at which the reduced density matrix relaxes at the infinite time limit \cite{Hoening2012,Horstmann2013}. The study of the steady state’s quantum properties as well as the rate at which this it is approached, is of great importance for both theoretical and practical reasons. A driven approach to the steady state has been used to deal with quantum information processing \cite{Kraus2008,Verstraete2009,Pastawski2011,Muschik2011}. The issue is standardly probed in the Lindbland master equation framework in which the steady state can be defined as the (right) eigenstate of the Lindblad super-operator with zero eigenvalues \cite{Prosen2008,Prosen2010}.

\par
In recent papers \cite{Lyris2021,Lyris2020}, we presented a scheme that combines the Feynman-Vernon’s influence functional technique with the out-of-equilibrium Keldysh formalism. This construction enables the direct calculation of the environment’s impact on the correlation functions of an open bosonic or fermionic quantum system. In this approach, the focus is not on the reduced density matrix but on the correlation functions per se. This permits the calculation of functions that contain more than a single time variable \cite{Lyris2020}. 

\par
In the present work, we apply this technique to probe the dynamics of quadratic 1D fermionic chains that when isolated, can be mapped on a spin chain model. The focus of the calculation is on the subsystem’s covariance matrix \cite{Horstmann2013}, the quantity that encaptures the properties of reduced density matrix. For the quadratic case, we find the steady state at which the system relaxes as well as the rate at which this state is approached. 

\par
The structure of the paper is the follows: In Section 2 we briefly present the basic ingredients of our formalism which is characterized by the the introduction of the reduced generating functional. The latter is written as a coherent state path integral over paths parametrized in terms of the Keldysh-Schwinger complex time variable. The environment is simulated by a collection of fermionic harmonic oscillators and, by assuming that it interacts linearly with the system, its degrees of freedom are integrated out. The resulting quantity, called influence functional, fully expresses the impact of the environment on the dynamics of the system. In this way, we construct the reduced generating functional through which the calculation of the covariance matrix, as well the calculation of any reduced correlation function, is immediate via functional differentiations. The results of this Section are quite general not depending on the specific system under consideration.

\par
In Section 3, we apply the aforementioned formalism for a system that is described by a Hamiltonian which is quadratic when written in terms of Majoranas variables and we derive the general form of the generating functional through which the all-important covariance matrix can be straightforwardly calculated. We also discuss the Markovian limit, on which the analytical calculations of the next Section are based.

\par
Section 4 refers to a fermionic system which, when isolated, is characterized by a quantum phase transition. In the framework of the present formalism, we examine its properties when it is part of a compound system. The site-translational symmetry it possesses, facilitates the diagonalization of the corresponding Hamiltonian providing, thus, a concrete example of the calculations presented in Section 3. We derive the covariance matrix and we examine its analytic properties at the steady state limit. The main result of this Section is the confirmation that the non-analyticity presented in the ground state correlations of the isolated system, remains in the steady state of the open system indicating the possible persistence of quantum correlations. 
\par
Finally, in Section 5 we present the conclusions of the current investigation. In Appendix A, we present some technical details that were left out in the main text.

\section{The Reduced Generating Functional}\label{reduced}
\setcounter{equation}{0}

We consider a compound fermionic system consisting of two parts: The (sub)system $S$ and its environment $E$. The dynamics of the isolated $S+E$ system is controlled by a Hamiltonian of the form: 

\begin{equation}
\label{2.1}
\hat{H}=\hat{H}_S\otimes \hat{I}_E+\hat{I}_S\otimes \hat{H}_E+\hat{H}_I
\end{equation}
The Hamiltonian operators in the last expression are defined in terms of fermionic creation and annihilation operators:

\begin{equation}
\label{2.2}
\hat{H}_S=\hat{H}\left( \hat{a}_{S}^{\dagger},\hat{a}_S \right),~~~\hat{H}_E=\hat{H}\left( \hat{a}_{E}^{\dagger},\hat{a}_E \right),~~~\hat{H}_I=\hat{H}\left( \hat{a}_{S}^{\dagger},\hat{a}_S;\hat{a}_{E}^{\dagger},\hat{a}_E \right) 
\end{equation}

The physical picture we adopt is the following: Up to a moment $t_{in.}$ the environment and the system are independent of each other $\left( \hat{H}_I\left( t\le t_{in.} \right) =0 \right)$ and the density operator of the compound system can be expressed in the product form $\hat{\rho}\left( t_{in.} \right) =\hat{\rho}_E\left( t_{in.} \right) \otimes \hat{\rho}_S\left( t_{in.} \right)$. Before the moment $t_{in.}$ the environment is in equilibrium at temperature $T=\beta ^{-1}$ meaning that $\hat{\rho}_E=\left( Z_E\left( \beta \right) \right) ^{-1}\exp \left( -\beta \hat{H}_E \right) $.  After the initialization of the interaction, the parts of the compound system entangle, and the reduced time evolution becomes non-unitary. 

As probes for the system’s dynamics, one usually considers correlations of the form:

\begin{equation}
\label{2.3}
G_{jk}\left( \rho ;t_2,t_1 \right) =Tr\left[ \hat{\rho}\left( t_{in.} \right) \hat{T}\left( \hat{a}_{S,j}^{\dagger}\left( t_2 \right) \hat{a}_{S,k}\left( t_1 \right) \right) \right]. 
\end{equation}

In the last expression, $\hat{a}_{S,j}\left( t \right) $ is a fermionic Heisenberg operator that refers to the system, the subscript $j$ is a site or space index, $\hat{\rho}\left( t_{in.} \right) $ is the initial density operator of the compound system, $\hat{T}$ is the time ordering operator and the trace operation refers to both the environment’s and the system’s degrees of freedom. 

Needless to say, the calculation of \eqref{2.3} and of any reduced correlation function, is not a trivial task. All the efforts, analytical or numerical, for confronting the issue are restricted in the framework of Lindblad's master equation \cite{Prosen2008,Prosen2010}. In the present study we shall adopt a more general formalism for the study of the system’s dynamics, based on the introduction of a functional that generates, via the application of the appropriate derivatives, the reduced functions we are interested in. 

To set the stage, we begin by noting that correlators of Heisenberg field operators are
characterized by a forward-backward time structure (see Eq. \eqref{2.3}). As long as the system is in equilibrium and isolated this structure is easily taken into account by assuming an adiabatic evolution of the non-interacting vacuum \cite{Peskin1995,Kamenev2009}. However, when the system is out of equilibrium or open, the adiabatic hypothesis is not valid. In that case the dynamics can only be analyzed  through the out-of-equilibrium Keldysh-Schwinger formalism. \cite{Keldysh1964,Keldysh1965}. In the present case, we examine a quantum system in contact with its surroundings or, put in other words, we focus on a (presumably small) part of an isolated compound system. As it is obvious, the situation shares a lot with the out of equilibrium dynamics. Taking this into account, we have presented in \cite{Lyris2021} a formalism based on a reduced version of the Keldysh-Schwinger correlation functions. It is based on an extension of Feynman-Vernon’s technique along the Keldysh time contour, an extension that focuses on the calculation of correlation functions bypassing the determination of the reduced density matrix per se. 

The scheme begins with the interpretation of the trace operation in Eq. \eqref{2.3} in the (over complete) coherent state basis $\left| \mathbf{z}_S \right> \equiv \otimes _{j\in S}\left| z_j \right> $ for the system and $\left| \bm{\upzeta}_E \right>  \equiv \otimes _{\mu \in E}\left| \zeta _{\mu} \right> $ for the environment, followed by the construction of the generating functional for system’s correlators.

Apart from the details about the peculiarities of defining path integration over
coherent states, the key ingredient in the aforementioned construction is that the
paths entering the generating functional are parametrized along the so-called Keldysh
time contour \cite{Keldysh1964,Keldysh1965}. 

The Keldysh time contour is defined on the complex plane along a closed contour $P$ that encircles the real $t$ axis running from $t_{in.+}\equiv t_{in.}+i0$ to $t_{in.-}\equiv t_{in.}-i0$. The contour consists of two straight lines. The first one, denoted as $L_+$, joins the point $t_{in.+}$ to an arbitrary time instance $T_+=T+i0$. Along this line the time variable is denoted as $t_+$. The second line joins $t_{-}=T-i0$ and along this line, defined as $L_-$, time is denoted as $t_-$. In the case of thermal initial states, the contour is extended by a complex time line, running parallel to the imaginary axis, from $t_{in.-}$ to $ t_{in.}-i\beta $, where $\beta ^{-1}$ is the temperature of the corresponding thermal state. This third line is denoted as $L_{\beta}$, while the extended contour is denoted as $C$.

\begin{figure}[!htb]
    \centering
    \includegraphics[scale=0.75]{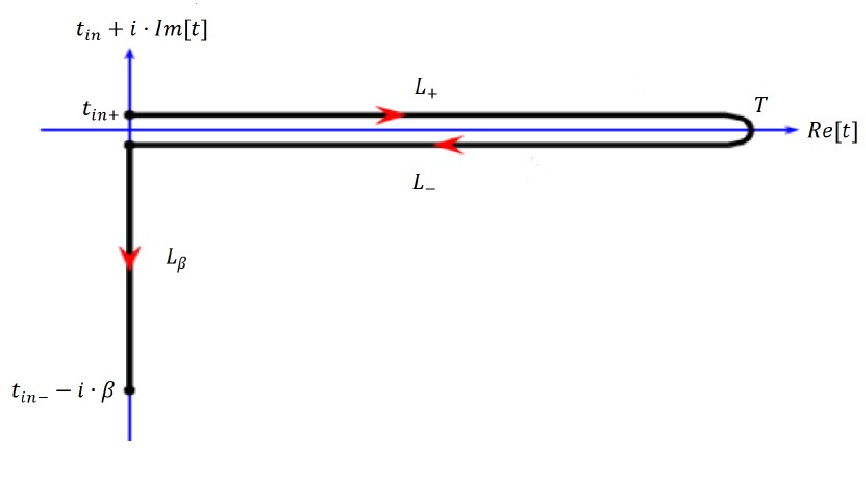}
    \caption{The Keldysh Contour}
    \label{fig:The_Keldysh_Contour}
  \end{figure} 

A natural ordering arises along this configuration since times along $L_+$ are considered to be before times along $L_-$, which are also considered to be before times along $L_{\beta}$ (see Fig. 1). Taking as granted the technical details presented in Refs. \cite{Lyris2021,Lyris2020}. we quote here the result for the reduced generating functional:

\begin{equation}
\begin{split}
\label{2.4}
Z_S\left[ \bar{\mathbf{{J}}},\mathbf{J} \right] & = \smallint d^2\mathbf{z}_Sd^2\mathbf{z}'_S\left< -\mathbf{z}_S \right|\hat{\rho}_S\left( t_{in.+} \right) \left| \mathbf{z}'_S \right> \\
& \times \int\limits_{\mathbf{z}\left( t_{in.+} \right) =\mathbf{z}_S}^{\bar{\mathbf{{z}}}\left( t_{in.-} \right) =\bar{\mathbf{{z}}}'_S}{D^2\mathbf{z}\left( t_P \right) e^{iS_S\left[ \bar{\mathbf{{z}}},\mathbf{z} \right] -E\left( \bar{\mathbf{{z}}}_S,\mathbf{z}_S \right) +i\int\limits_P{dt_P\left( \mathbf{\bar{J}}\cdot \mathbf{\dot{z}}+\mathbf{\bar{z}}\cdot \mathbf{\dot{J}} \right) }}I_E\left[ \bar{\mathbf{{z}}},\mathbf{z} \right]}.
\end{split}
\end{equation}

At this point explanations are needed for the last expression to become transparent: The Grassmann fields $\mathbf{z}\left( t_P \right) \equiv \mathbf{z}_P\left( t \right) $ entering in the path integral are defined along the Keldysh contour $P$ (assuming that the system is at zero temperature):

\begin{equation}
\label{2.5}
\mathbf{z}\left( t_P \right) =\left\{ \begin{array}{c}
	\mathbf{z}\left( t_+ \right) \equiv \mathbf{z}_+~~\mathrm{ along}~~L_+\\
	\mathbf{z}\left( t_- \right) \equiv \mathbf{z}_-~~\mathrm{ along}~~L_-\\
\end{array} \right. .
\end{equation}

These variables are considered as independent and they are integrated separately. The action $S$ refers to the system and assumes the following form:

\begin{equation}
\label{2.6}
S_S=\int\limits_P{dt_P}\left[ \frac{i}{2}\left( \mathbf{\bar{z}}\cdot \mathbf{\dot{z}}-\mathbf{\dot{\bar{z}}}\cdot \mathbf{z} \right) -H_S\left( \mathbf{\bar{z}},\mathbf{z} \right) \right].
\end{equation}

In Eq. \eqref{2.4} the action is accompanied by the surface term
\begin{equation}
\label{2.7}
E\left( \mathbf{\bar{z}}'_S,\mathbf{z}_S \right) =\frac{1}{2}\left[ \left| \mathbf{\bar{z}}'_S \right|^2+\left| \mathbf{z}_S \right|^2 \right] -\frac{1}{2}\left[ \mathbf{\bar{z}}'_S\cdot \mathbf{z}\left( t_{in.-} \right) +\mathbf{\bar{z}}\left( t_{in.+} \right) \cdot \mathbf{z}_S \right]. 
\end{equation}

The generating functional depends on the initial state of the system through the factor $\left< -\mathbf{z}_S \right|\hat{\rho}_S\left( t_{in.+} \right) \left| \mathbf{z}'_S \right> $. When the system is initially in its ground state, $\hat{\rho}_S\left( t_{in.+} \right)=|GS\rangle \langle GS|$,  this factor together with the surface term are integrated out and the generating functional reduces to the simpler form:  

\begin{equation}
\label{2.8}
Z_{S}^{\left( 0 \right)}\left[ \mathbf{\bar{J}},\mathbf{J} \right] =\int{D^2\mathbf{z}\left( t_P \right) e^{iS_S\left[ \mathbf{\bar{z}},\mathbf{z} \right] +i\int\limits_P{dt_P\left( \mathbf{\bar{J}}\cdot \mathbf{z}+\mathbf{\bar{z}}\cdot \mathbf{J} \right)}}I_E\left[ \mathbf{\bar{z}},\mathbf{z} \right]}.
\end{equation}

Sources have been added in Eq. \eqref{2.4} to produce the system’s correlation functions through functional differentiation. For example, the correlation function \eqref{2.3} can be derived by using the formula:

\begin{equation}
\label{2.9}
G_{jk}\left( \rho ;t_2,t_1 \right) =\left. \frac{\delta ^2\ln Z_{S}^{\left( 0 \right)}}{\delta J_j\left( t_{1+} \right) \delta \bar{J}_k\left( t_{2+} \right)} \right|_{J=0}.
\end{equation}

The functional $I_E$, appearing in Eq. \eqref{2.4} is produced after tracing out the environmental degrees of freedom. It is the reason we consider $Z_S$ as reduced. It incorporates the influence of the environment on the system’s dynamics and has the same significance as the Feynman-Vernon’s influence functional for the calculation of the reduced density matrix. It is defined as

\begin{equation}
\label{2.10}
I_E\left[ \mathbf{\bar{z}},\mathbf{z} \right] =\frac{1}{Z_E\left( \beta \right)}\int\limits_{AP}{D\bm{\upzeta }\left( t_C \right) e^{i\int\limits_C{dt_C\left( \mathbf{\bar{\bm{\upzeta}}}\cdot \mathbf{\dot{\bm{\upzeta}}}-H_E\left( \mathbf{\bar{\bm{\upzeta}}},\bm{\upzeta } \right) \right) -i\int\limits_C{dt_CH_I\left( \mathbf{\bar{\bm{\upzeta}}},\bm{\upzeta };\mathbf{\bar{z}},\mathbf{z} \right)}}}}.
\end{equation}

Here, the paths are parametrized along the full Keldysh contour $C=P\cup L_{\beta}$ with the agreement that interactions are absent along the thermal branch $L_{\beta}$. The antiperiodic boundary conditions are induced by the fermionic trace operation. As expected, the calculation of the influence functional is not a trivial task. In the present work, we adopt the simplifying assumption that the environment can be simulated by a collection of (fermionic) harmonic oscillators which interact linearly with the system:

\begin{equation}
\begin{split}
\label{2.11}
\hat{H}_E =\sum_{\mu}{E_{\mu}}\left( \hat{a}_{E,\mu}^{\dagger}\hat{a}_{E,\mu}-\frac{1}{2} \right),  ~~~ \hat{H}_I =\sum_{j\in S,\mu \in E}{\left( \lambda_{j\mu}\hat{a}_{E,\mu}^{\dagger}\hat{a}_{S,j}+\bar{\lambda}_{j\mu}\hat{a}_{S,j}^{\dagger}\hat{a}_{E,\mu} \right)}.
\end{split}
\end{equation}

This interaction results in fermion number non-conservation and is responsible for the dissipation effects we are interested in. In this case, the influence functional can be exactly calculated \cite{Lyris2020}:

\begin{equation}
\label{2.12}
I_E=C_{E}^{-1}\exp \left[ -\int\limits_P{dt_P\int\limits_{P'}{dt'_{P'}\bar{\mathbf{z}}\left( t_P \right) \mathbf{\Delta } \left( t_P-t'_{P'} \right) \mathbf{z}\left( t'_{P'} \right)}} \right].
\end{equation}

The different indexing in time integration indicates the fact that the time parameters in the last expression can run along different branches of the Keldysh contour. The constant appearing in Eq. \eqref{2.12}, $C_{E}^{-1}=\prod_{\mu \in E}{2\cosh \left( \beta E{{_{\mu}}/{2}} \right)}$ , is of no significance for the calculation of correlation functions. The crucial quantities are the matrix elements $\left( \mathbf{\Delta } \right) _{jk}=\Delta _{jk}$ which assume the form \cite{Lyris2020}:

\begin{equation}
\begin{split}
\label{2.13}
\Delta _{jk}\left( t_P-t'_{P'} \right) & =\sum_{\mu \in E}{\bar{\lambda}_{j\mu}\lambda_{k\mu}}\left[ \Theta \left( t_P-t'_{P'} \right) -\frac{1}{1+e^{\beta E_{\mu}}} \right] e^{-i\left( t_P-t'_{P'} \right) E_{\mu}}= \cr
& = \int\limits_0^{\infty}{dED_{jk}\left( E \right)}\left[ \Theta \left( t_P-t'_{P'} \right) -\frac{1}{1+e^{\beta E}} \right] e^{-i\left( t_P-t'_{P'} \right) E}.
\end{split}
\end{equation}

Here we defined:

\begin{equation}
\label{2.14}
D_{jk}\left( E \right) =\sum_{\mu \in E}{\bar{\lambda}_{j\mu}}\lambda_{k\mu}\delta \left( E-E_{\mu} \right).
\end{equation}

The most general form of the system-environment interaction \cite{Breuer2007} is usually written in the form $\sum\limits_\alpha  {{{{\rm{\hat S}}}_\alpha } \otimes } {\rm{ }}{\hat {\rm E}_\alpha }$ where ${{\rm{\hat S}}_\alpha }$ and ${\hat {\rm E}_\alpha }$ are operators acting on the system and the environment respectively. In the present case, this means that each site of the environment interacts with the linear combination $\hat a_{S,\mu }^{eff} = \sum\limits_{j \in S} {{\lambda _{j\mu }}} {\hat a_{S,j}}$, a fact that induces extra interactions in the embedded system. Equivalently and without losing generality, this can be taken into account by writing:
\begin{equation}
\label{2.15}
{\bar \lambda _{j\mu }}{\lambda _{k\mu }} = {g_{jk}}{\left| {{\gamma _\mu }} \right|^2},{\rm{ }}{g_{jk}} = {\bar g_{kj}}
\end{equation}

As it is obvious the coefficients $g_{jk}$ characterize the strength of the interaction between the sites $j,k$ of the system, induced by the environment. The two extreme choices $g_{jk}=g\delta_{jk}$ and $g_{jk}=g~~\forall j,k$ correspond, the first one, to the case where each site of the system interacts independently with the environment, and the second one to the case in which all the sites of the system interact with each other with the same strength, irrespective of their distance. 

Thus, the influence of the environment, as it is encoded in Eq. \eqref{2.12}, yields a contribution to the generating functional which, in general, is nonlocal both in space and time indices. Inserting Eq. \eqref{2.12} into Eq. \eqref{2.8} we find the expression that we indent to use for the calculation in the current investigation:

\begin{equation}
\label{2.16}
Z_{S}^{\left( 0 \right)}\left[ \mathbf{\bar{J}},\mathbf{J} \right] =C_{E}^{-1}\int{D^2\mathbf{z}\left( t_P \right) e^{iS_S\left[ \mathbf{\bar{z}},\mathbf{z} \right] +iS_{INF.}\left[ \mathbf{\bar{z}},\mathbf{z} \right] +i\int\limits_P{dt_P\left( \mathbf{\bar{J}}\cdot \mathbf{z}+\mathbf{\bar{z}}\cdot \mathbf{J} \right)}}}
\end{equation}
with
\begin{equation}
\label{2.17}
S_{INF.}=i\int\limits_P{dt_P\int\limits_{P^{\mathrm{'}}}{dt'_{P'}}}\mathbf{\bar{z}}\left( t_P \right) \mathbf{\Delta }\left( t_P-t'_{P'} \right) \mathbf{z}\left( t'_{P'} \right).
\end{equation}

The generating functional in Eq. \eqref{2.16} can, in principle at least, be exactly calculated if the system’s Hamiltonian is quadratic.  In such a case the covariance matrix can also be exactly calculated. We shall confront the issue in the next chapter.  At this point, is worth note that through the (reduced) generating functional correlations of the form $\left< \hat{O}_{1,S}\left( t_1 \right) \hat{O}_{2,S}\left( t_2 \right) ...\hat{O}_{n,S}\left( t_n \right) \right> _{\rho}$  can be calculated. This must be contrasted to the case of calculations based on Lindblad’s equation where one considers correlations that contain a single time variable \cite{Breuer2007}. 

\section{Quadratic Hamiltonians}
\label{quadratic}
\setcounter{equation}{0}

In this chapter we are interested in systems the dynamics of which are based on quadratic fermionic Hamiltonians. It would be convenient to rewrite them in terms of the real Majorana fields not only because these are the basis for defining fermionic path integration \cite{Lyris2020} but also because it is the standard stage for diagonalizing the relevant Hamiltonians \cite{Osborne2002,Vidal2003}. In any case, in the present work, the physical degrees of freedom are the fermionic ones and the Majorana variables constitute a convenient change of variables performed through a canonical transformation. With this analysis in mind, we write:

\begin{equation}
\label{3.1}
H_S=\frac{i}{4}\sum_{j,k=1}^N{\sum_{u,v=0,1}{\gamma _{j,u}}}A_{ju,kv}\gamma _{k,v}=\frac{i}{4}\sum_{\alpha ,\beta =1}^{2N}{\gamma _{\alpha}}A_{\alpha \beta}\gamma _{\beta}\equiv \frac{i}{4}\bm{\upgamma}\mathbf{A}\bm{\upgamma}.
\end{equation}

The Majorana variables entering the last expression are defined according to the rule:

\begin{equation}
\label{3.2}
\gamma _{2j-1}=\bar{z}_{S,j}+z_{S,j}\equiv \gamma _{j,0},~~~\gamma _{2j}=i\left( \bar{z}_{S,j}-z_{S,j} \right) \equiv \gamma _{j,1}.
\end{equation}

In Eq. \eqref{3.1} we adopted the compact notation of Ref. \cite{Horstmann2013} $\alpha=\left( j,u \right)$ in which $j=1,...,N$ and $u=0,1$. In accordance, in the present Section, vectors and matrices are defined in a $2N$ dimensional space:

\begin{equation}
\label{3.3}
\bm{\upgamma}\equiv \left( \gamma _1,\gamma _2,...,\gamma _{2N-1},\gamma _{2N} \right) =\left( \gamma _{1,0},\gamma _{1,1},...,\gamma _{N,0},\gamma _{N,1} \right).
\end{equation}

The $2N\times 2N$ antisymmetric matrix $\mathbf{A}\doteq A_{\alpha \beta}=A_{jk,uv}=-A_{kj,vu}=-A_{\beta \alpha}$ describes the interaction between the sites of the system. In the quadratic case, the generating functional \eqref{2.16} can be exactly calculated by minimizing the action:

\begin{equation}
\label{3.4}
S=S_S+S_{INF.}+S_J.
\end{equation}

Due to form \eqref{3.1} of the Hamiltonian, it is convenient to express $S$ in terms of Majorana variables:

\begin{equation}
\begin{split}
\label{3.5}
S & =  \frac{i}{4}\int\limits_P{dt_P}\bm{\upgamma}\left( t_P \right) \left( \mathbf{I}\partial _{t_P}-\mathbf{A} \right) \bm{\upgamma}\left( t_P \right)  +\frac{i}{4}\int\limits_P{dt_P}\int\limits_{P'}{dt'_{P'}}\bm{\upgamma}\left( t_P \right) \mathbf{D}\left( t_P,t'_{P'} \right) \bm{\upgamma}\left( t'_{P'} \right) \\ & +\frac{1}{2}\int\limits_P{dt_P}\mathbf{f}\left( t_P \right) \cdot  \bm{\upgamma}\left( t_P \right) .
\end{split}
\end{equation}

In the last expression we introduced the sources:

\begin{equation}
\label{3.6}
f_{j,0}=\bar{J}_j-J_j,~~~f_{j,1}=i\left( \bar{J}_j+J_j \right) 
\end{equation}
and we defined
\begin{equation}
\label{3.7}
\mathbf{D}\doteq D_{\alpha \beta}=D_{ju,kv}=\Delta _{jk}\left( 1-\sigma ^y \right) _{uv}.
\end{equation}

By minimizing Eq. \eqref{3.5} we get the following ``classical'' equation:
\begin{equation}
\label{3.8}
\left( \mathbf{I}\partial _{t_P}-\mathbf{A} \right) \bm{\upgamma}^{cl.}\left( t_P \right) +\int\limits_{P'}{dt'_{P'}}\tilde{D}\left( t_P,t'_{P'} \right) \bm{\upgamma}^{cl.}\left( t'_{P'} \right) =-i\mathbf{f}\left( t_P \right) 
\end{equation}
in which
\begin{equation}
\label{3.9}
\tilde{D}_{\alpha \beta}\left( t_P,t'_{P'} \right) =\frac{1}{2}\left( D_{\alpha \beta}\left( t_P,t'_{P'} \right) -D_{\beta \alpha}\left( t'_{P'},t_P \right) \right) =-\tilde{D}_{\beta \alpha}\left( t_P,t'_{P'} \right) .
\end{equation}

By introducing Green’s function $\Gamma_{\alpha \beta}$ as the causal solution of Green’s equation   

\begin{equation}
\label{3.10}
\int\limits_{P''}{dt''_{P''}}\left[ \left( \mathbf{I}\partial _{t_P}-\mathbf{A} \right) \delta \left( t_P-t''_{P''} \right) + \mathbf{\tilde{D}}\left( t_P,t''_{P''} \right) \right] \mathbf{\Gamma }\left( t''_{P''},t'_{P'} \right)  =-\mathbf{I}i\delta \left( t_P-t'_{P'} \right). 
\end{equation}
the generating functional \eqref{2.16} assumes the form:

\begin{equation}
\label{3.11}
Z_{S}^{\left( 0 \right)}\left[ \mathbf{f} \right] =C_{E}^{-1}\exp \left[ \frac{i}{4}\int\limits_{P'}{dt'_{P'}\int\limits_P{dt_P\mathbf{f}\left( t_P \right) \mathbf{\Gamma }\left( t_P,t'_{P'} \right) \mathbf{f}\left( t'_{P'} \right)}} \right] .
\end{equation}

As it is obvious, having in hand the solution of Eq. \eqref{3.10} and the generating functional \eqref{3.11}, one can calculate, via a unitary transformation of the sources, all the correlation functions of the system. 

In the present study, we focus on the real and antisymmetric covariance matrix that is defined as \cite{Horstmann2013}:

\begin{equation}
\label{3.12}
C_{jk,uv}\left( t \right) =\mathrm{Tr}\left\{ \hat{\rho}\left( t \right) \frac{i}{2}\left[ \hat{\gamma}_{j,u},\hat{\gamma}_{k,v} \right] \right\} =\mathrm{Tr}_{\mathrm{S}}\left\{ \hat{\rho}_{S}^{R}\left( t \right) \frac{i}{2}\left[ \hat{\gamma}_{j,u},\hat{\gamma}_{k,v} \right] \right\} .
\end{equation}

The key ingredient in this expression is the reduced density matrix

\begin{equation}
\label{3.13}
\hat{\rho}_{S}^{R}\left( t \right) =Tr_E\left\{ e^{-i\left( t-t_{in.} \right) \hat{H}}\hat{\rho}\left( t_{in.} \right) e^{i\left( t-t_{in.} \right) \hat{H}} \right\} 
\end{equation}
which, in the present case, is a Gaussian state and, as such, all its properties are encoded into the covariance matrix \eqref{3.12}.

Knowledge of the covariance matrix permits physically important issues to be addressed. One of the most challenging refers to the rate at which correlations diminish due to decoherence. Closely related to it is the approach of the reduced density matrix to the, so-called, steady state, $\hat{\rho}_{S}^{R}\left( t \right) \underset{t\rightarrow \infty}{\rightarrow}\hat{\rho}_{S,0}^{R}$ and the consequent approach of the covariance matrix to the corresponding steady state covariance: $C_{jk,uv}\left( t \right) \underset{t\rightarrow \infty}{\rightarrow}C_{jk,uv}^{\left( 0 \right)}$.  At the thermodynamic limit $N\rightarrow \infty $, the knowledge of the analytic structure of these functions is of great importance as it can reveal possible quantum phase transitions. 

The covariance matrix can be straightforwardly calculated through the generating functional \eqref{3.11} by applying functional derivatives as follows:
\begin{equation}
\label{3.14}
C_{jk,uv}\left( t \right) =-4i\left. \frac{\delta ^2\ln Z_{S}^{\left( 0 \right)}}{\delta f_{ju}\left( t_+ \right) \delta f_{kv}\left( t_+ \right)} \right|_{J=0}.
\end{equation}

In the present approach, the key function for achieving the final result is the solution of  Green’s equation \eqref{3.10}.  However, this equation, in its general form, is nonlocal both in space and time indices, a fact that makes its analytical solution a very difficult task. Even so, it can be written in a way more convenient for calculations as the real antisymmetric matrix $A_{\alpha \beta}$ can always be diagonalized through an orthogonal transformation:
\begin{equation}
\label{3.15}
\mathbf{Q}^{\bot}\mathbf{AQ}=\oplus _{j=1}^{N} \left(
\begin{array}{cc}
0 & \varepsilon _j \\
-\varepsilon _j & 0 \\
\end{array}
\right) ,~~\varepsilon _j\ge 0,~~\mathbf{Q}\in O\left( 2N \right) .
\end{equation}

Thus, by using the unitary transformation 

\begin{equation}
\label{3.16}
\mathbf{V}=\mathbf{Q}\oplus ^N\frac{1}{\sqrt{2}} \left(
\begin{array}{cc}
1 & 1 \\
i & -i \\
\end{array}
\right) 
\end{equation}
we can write

\begin{equation}
\label{3.17}
\mathbf{V}^{\dagger}\mathbf{AV}=\oplus ^N \left(
\begin{array}{cc}
i\varepsilon _j & 0 \\
0 & -i\varepsilon _j \\
\end{array}
\right) \equiv i\mathbf{E}.
\end{equation}

By writing
\begin{equation}
\begin{split}
\label{3.18}
& \sum_{j,u}{\gamma _{j,u}}V_{jk,uv}=\psi _{k,v}\rightarrow \bm{\upgamma} \cdot \mathbf{V}=\bm{\uppsi},~~\mathbf{K}=\mathbf{V}^{\dagger}\mathbf{\tilde{D}V}, 
\\ & \mathbf{j}=\mathbf{f}\cdot \mathbf{V}\rightarrow j_{k,v}=\sum_{j,u}{f_{j,u}}V_{jk,uv}
\end{split}
\end{equation}
the action \eqref{3.5} can be written as follows:

\begin{equation}
\begin{split}
\label{3.19}
S&=\frac{i}{4}\int\limits_P{dt_P}\bm{\uppsi}\left( t_P \right) \left( \mathbf{I}\partial _{t_P}-i\mathbf{E} \right) \bm{\uppsi}\left( t_P \right) \\ & +\frac{i}{4}\int\limits_P{dt_P}\int\limits_{P'}{dt'_{P'}}\bm{\uppsi}\left( t_P \right) \mathbf{K}\left( t_P,t'_{P'} \right) \bm{\uppsi}\left( t'_{P'} \right) \\ & -\frac{1}{4}\int\limits_P{dt_P}\mathbf{\bar{j}}\left( t_P \right) \cdot \bm{\uppsi}\left( t_P \right) -\frac{1}{4}\int\limits_P{dt_P}\bar{\bm{\uppsi}}\left( t_P \right) \cdot \mathbf{j}\left( t_P \right) .
\end{split}
\end{equation}

In this case the classical equation of motion assumes the form:

\begin{equation}
\label{3.20}
\left( \mathbf{I}\partial _{t_P}-i\mathbf{E} \right) \bm{\uppsi}^{\mathbf{cl}.}\left( t_P \right) +\int\limits_{P'}{{dt'}_{P'}}\mathbf{K}\left( t_P,t'_{P'} \right) \bm{\uppsi}^{\mathbf{cl}.}\left( t'_{P'} \right) =-i\mathbf{j}\left( t_P \right) .
\end{equation}

Thus, by defining the function $\mathbf{L}=\mathbf{V}^{\mathbf{\dagger }}\mathbf{\Gamma V}$, the Green’s equation \eqref{3.10} is recasted as:

\begin{equation}
\label{3.21}
\left( \mathbf{I}\partial _{t_P}-i\mathbf{E} \right) \mathbf{L}\left( t_P,t'_{P'} \right) +\int\limits_{P''}{{dt''}_{P''}}\mathbf{K}\left( t_P,{t''}_{P''} \right) \mathbf{L}\left( {t''}_{P''},t'_{P'} \right) =-\mathbf{I}i\delta \left( t_P-t'_{P'} \right) 
\end{equation}

The last expression is a set of equations coupled due to the presence of the kernel $\mathbf{K}$, the dissipation kernel. In the current paper we adopt an approximation that can lead to the exact solution of the Green’s Eqs. \eqref{3.21}. It is the same approximation on which the Lindbland approach is founded namely the, so-called, Markovian approximation \cite{Breuer2007}. It is based on the hypothesis that the environment acts as a memoryless bath and, consequently, that the influence functional can be considered as local in time. The validity of such an approximation is based on the existence of two characteristic time scales \cite{Preskill2015}. The first one, $\tau_E$, refers to the decay of environmental correlations, while the second one, $\tau _{S}$, characterizes the frequencies $\varepsilon _S \sim \tau _{S}^{-1}$ after which system’s dynamics are screened out. If the last scale is much larger than the first one, the function \eqref{3.13} behaves much like as a delta function:
\begin{equation}
\label{3.22}
\Delta _{jk}\left( t_P-t'_{p'} \right) \rightarrow g_{jk}\Delta _{PP'}\left( \varepsilon _S \right) \delta \left( t-t' \right) 
\end{equation}

The matrix elements $\Delta _{PP'}$  are defined as \cite{Lyris2020}:

\begin{equation}
\begin{split}
\label{3.23}
\Delta _{++} =-i\delta E+\Gamma \left( \frac{1}{2}-b \right) ,~\Delta _{--} =i\delta E+\Gamma \left( \frac{1}{2}-b \right) ,
~\Delta _{+-}   =-\Gamma b,~
\Delta _{-+}  =\Gamma \left( 1-b \right)
\end{split}
\end{equation}
with:

\begin{equation}
\label{3.23}
\Gamma =2\pi D\left( \varepsilon _S \right),~~\delta E=\mathrm{Pr}.\int\limits_0^{\infty}{dE}\frac{D\left( E \right)}{E-\varepsilon _S},~~b=\frac{1}{1+e^{\beta \varepsilon _S}}
\end{equation}

The Markovian limit makes the dissipation kernel  diagonal in the time variables. Even so, expression \eqref{3.21} is a set of $4\times \left( 2N \right) ^ 2$ coupled equations the solution of which is a rather complicated task. However, when the Hamiltonian is translationally invariant Eqs. \eqref{3.21} can be exactly solved. This is the task of the next Section.

\section{Kitaev’ s chains}
\label{chain}
\setcounter{equation}{0}

In the present Section we examine a simple -yet interesting- fermionic model in one spatial dimension, namely the $1D$ superconducting chain \cite{Osborne2002,Kitaev2009,Wei2011}:

\begin{equation}
\begin{split}
\label{4.1}
\hat{H}_S=\sum_{j=1}^N{\left[ -w\left( \hat{a}_{j}^{\dagger}\hat{a}_{j+1}+\hat{a}_{j+1}^{\dagger}\hat{a}_j \right) +\Delta \left( \hat{a}_j\hat{a}_{j+1}+\hat{a}_{j+1}^{\dagger}\hat{a}_{j}^{\dagger} \right) -\mu \left( \hat{a}_{j}^{\dagger}\hat{a}_j-\frac{1}{2} \right) \right]}
\end{split}
\end{equation}

Although simple and completely solvable when isolated, this model is quite rich due to the underlying quantum phase transition. After the Jordan-Wigner transformation

\begin{equation}
\label{4.2}
\hat{a}_j=\prod_{k=1}^{j-1}{\left( -\sigma _{k}^{z} \right)}\sigma _{j}^{-},~~~\hat{a}_{j}^{\dagger}=\prod_{k=1}^{j-1}{\left( -\sigma _{k}^{z} \right)}\sigma _{j}^{+},~~~\sigma _{j}^{\pm}=\frac{1}{2}\left( \sigma _{j}^{x}\pm i\sigma _{j}^{y} \right) 
\end{equation}
the fermionic system is mapped onto the anisotropic $XY$ spin system:

\begin{equation}
\label{4.3}
\hat{H}_S=\sum_j{\left( \frac{\Delta -w}{2}\hat{\sigma}_{j}^{x}\hat{\sigma}_{j+1}^{x}+\frac{\Delta +w}{2}\hat{\sigma}_{j}^{y}\hat{\sigma}_{j+1}^{y}-\frac{\mu}{2}\hat{\sigma}_{j}^{z} \right)}.
\end{equation}

For reasons of simplicity, we choose $w=-\Delta=1$ and we write $\mu=2h$. In this case, the spin system \eqref{4.3} reduces to the transverse Ising model:

\begin{equation}
\label{4.4}
\hat{H}_S=-\sum_j{\left( \hat{\sigma}_{j}^{x}\hat{\sigma}_{j+1}^{x}+h\hat{\sigma}_{j}^{z} \right)}
\end{equation}
while its fermionic ancestor reads:
\begin{equation}
\label{4.5}
\hat{H}_S=-\sum_{j=1}^N{\left[ \left( \hat{a}_{j}^{\dagger}-\hat{a}_j \right) \left( \hat{a}_{j+1}^{\dagger}+\hat{a}_{j+1} \right) +2h\left( \hat{a}_{j}^{\dagger}\hat{a}_j-\frac{1}{2} \right) \right]}.
\end{equation}

In the presence of the environment, the dynamics of these systems become quite complicated studied mainly in the framework of the Lindbland master equation. From this point of view the model \eqref{4.1}, and all its integrable variants, provides the ideal stage for presenting the analytic formalism introduced in Sections 2 and 3.  However, a note is needed at this point:  When the system we are interested in is isolated, the Hamiltonians \eqref{4.1} and \eqref{4.3} are mathematically equivalent meaning that the calculation of the relevant correlation functions can be based either on the first or the second one. When the system is part of a larger compound system, it is the mutual interaction that defines the physical degrees of freedom. In the case under study the environment is a fermionic bath meaning that the physical degrees of freedom for the description of the system are fermionic and the relevant quantum Hamiltonian is \eqref{4.3}.

The classical Hamiltonian representing the fermionic system \eqref{4.5} in the integral \eqref{2.7}, can be written in the form (see Eq. \eqref{3.2}): 
\begin{equation}
\begin{split}
\label{4.6}
H_S&=\sum_j{\left[ \left( z_jz_{j+1}-\bar{z}_j\bar{z}_{j+1} \right) +\left( z_j\bar{z}_{j+1}-\bar{z}_jz_{j+1} \right) -2h\bar{z}_jz_j \right]}=\\ & =\frac{i}{4}\sum_{jk,uv}{\gamma _{j,u}A_{jk,uv}\gamma _{k,v}}
\end{split}
\end{equation}

As noted in the introduction of Section 3 it would be convenient to use the canonical transformation \eqref{3.2} to write the system's Hamiltonian in terms of Majorana variables.

The matrix $A$ entering the Majorana expression of the Hamiltonian is defined as:

\begin{equation}
\begin{split}
\label{4.7}
&A_{j,j+1;0,0}=-A_{j+1,j;0,0}=2,A_{j,j+1;0,1}=-A_{j+1,j;1,0}=-2, \\& A_{j,j;0,1}=-A_{j,j;1,0}=2h.
\end{split}
\end{equation}

The first step for obtaining the solution of the classical equation \eqref{3.10} is the diagonalization of system’s Hamiltonian. For the isolated case, this task is accomplished via the implementation of the discrete Fourier transform \cite{Keldysh1964,Keldysh1965,Wei2011}:

\begin{equation}
\label{4.8}
z_j=\frac{1}{\sqrt{N}}\sum_{m=0}^{N-1}{e^{i\varphi _mj}}c_m .
\end{equation}

The exact form of the phase $\varphi _m=\frac{2\pi}{N}\left( m+\kappa \right) $ depends on the number of fermions in the system. When this number is even, $\kappa=1/2$, while $\kappa=0$ when odd. The first case is connected with the antiperiodic boundary condition $z_{j+N}=-z_j$ and the latter with the periodic one $z_{j+N}=z_j$ \cite{Wei2011}. When dissipation is present, the fermion number is not conserved meaning that, strictly speaking, the transformation \eqref{4.8} should not be used.  However, at the thermodynamic limit $N\rightarrow \infty $ which we are interested in, the summation in Eq. \eqref{4.6} is extended over $\mathbb{Z} $ and the discrete transform \eqref{4.8} can be replaced by its continuum version:

\begin{equation}
\label{4.9}
z_j=\frac{1}{\sqrt{2\pi}}\int\limits_{-\pi}^{\pi}{d\varphi}e^{i\varphi j}c_{\varphi},~~~c_{\varphi}=\frac{1}{\sqrt{2\pi}}\sum_{j=-\infty}^{\infty}{e^{-i\varphi j}}z_j.
\end{equation}

These transformations are going to be helpful only if the system-environment interaction does not destroy the translational invariance presented in the isolated system. To preserve this symmetry, we shall assume that $g_{jk}=g_{j-k}=g_{k-j}$.

For the Majorana variables, the Fourier transforms read as follows:

\begin{equation}
\label{4.10}
\gamma _{j,u}=\frac{1}{\sqrt{2\pi}}\int\limits_{-\pi}^{\pi}{d\varphi}e^{i\varphi j}\gamma _{\varphi ,u},~~~\gamma _{\varphi ,0}=\bar{c}_{-\varphi}+c_{\varphi},~~~\gamma _{\varphi ,1}=i\left( \bar{c}_{-\varphi}-c_{\varphi}\right) .
\end{equation}

Inserting expressions \eqref{4.10} into Eq. \eqref{4.6}, the Hamiltonian can be rewritten in terms of the conjugate fields in the following form:

\begin{equation}
\label{4.11}
H_S=\int\limits_0^{\pi}{d\varphi}\bar{\bm{\upgamma}}_{\varphi}\mathbf{A}_{\varphi}\bm{\upgamma} _{\varphi}.
\end{equation}

To arrive at the last expression, we used the translational invariance of the Hamiltonian to write $A_{jk,uv}=A_{j-k,uv}$, we defined the vectors $\bm{\upgamma} _{\varphi}=\left( \gamma _{\varphi ,0},\gamma _{\varphi ,1} \right) $ and the matrices:

\begin{equation}
\label{4.12}
\left( \mathbf{A}_{\varphi} \right) _{uv}\equiv A_{\varphi ,uv}=\sum_j{e^{-i\varphi j}}A_{j,uv}=-A_{-\varphi ,vu},A_{j,uv}=\frac{1}{2\pi}\int\limits_{-\pi}^{\pi}{d\varphi e^{i\varphi j}}A_{\varphi ,uv}.
\end{equation}

It’s not difficult to find that:

\begin{equation}
\label{4.13}
\mathbf{A}_{\varphi}=-\frac{i}{2}\left( \begin{array}{cc}
	0&		h+e^{-i\varphi}\\
	-\left( h+e^{i\varphi} \right)&		0\\
\end{array} \right) .
\end{equation}

By using the unitary transformation

\begin{equation}
\label{4.14}
\mathbf{V}_{\varphi}=\frac{1}{\sqrt{2}}\left( \begin{array}{cc}
	1&		1\\
	-i&		i\\
\end{array} \right) e^{-i\theta _{\varphi}\sigma ^x}
\end{equation}
with
\begin{equation}
\label{4.15}
\cos 2\theta _{\varphi}=\frac{h+\cos \varphi}{\varepsilon _{\varphi}},~~\sin 2\theta _{\varphi}=\frac{\sin \varphi}{\varepsilon _{\varphi}},~~\varepsilon _{\varphi}=\sqrt{\left( h+\cos \varphi \right) ^2+\sin ^2\varphi}
\end{equation}
the matrix \eqref{4.13} can be put into diagonal form:

\begin{equation}
\label{4.16}
\mathbf{A}_{\varphi}=-\frac{1}{2}\mathbf{V}_{\varphi}\left( \begin{array}{cc}
	\varepsilon _{\varphi}&		0\\
	0&		-\varepsilon _{\varphi}\\
\end{array} \right) \mathbf{V}_{\varphi}^{\dagger}.
\end{equation}

By defining

\begin{equation}
\label{4.17}
\bm{\uppsi}_{\varphi}={\mathbf{V}^{\mathbf{\dagger }}}_{\varphi}\bm{\upgamma}_{\varphi},~~~\bar{\bm{\uppsi}}_{\varphi}=\mathbf{\bar{\bm{\upgamma}}}_{\varphi}\mathbf{V}_{\varphi}
\end{equation}
the Hamiltonian \eqref{4.11} simplifies as follows:
\begin{equation}
\label{4.18}
H_S=-\frac{1}{2}\int\limits_0^{\pi}{d\varphi \varepsilon _{\varphi}}{\bar{\bm{\uppsi}}}_{\varphi}\sigma ^z\bm{\uppsi}_{\varphi}.
\end{equation}

Introducing the function 
\begin{equation}
\label{4.19}
\tilde{g}_{\varphi}=\frac{1}{\sqrt{2\pi}}\sum_j{e^{i\varphi j}}g_j=\tilde{g}_{-\varphi}
\end{equation}
the influence of the environment is written as

\begin{equation}
\label{4.20}
S_{INF.}=\frac{i}{2}\int\limits_0^{\pi}{d\varphi}\int\limits_P{dt_P}\int\limits_{P^{\mathrm{'}}}{dt'_{P'}}{\bar{\bm{\uppsi}}}_{\varphi}\left( t_P \right) \mathbf{K}_{\varphi}\left( t_P,t'_{P'} \right) \bm{\uppsi}_{\varphi}\left( t'_{P'} \right) 
\end{equation}
where
\begin{equation}
\label{4.21}
\mathbf{K}_{\varphi}=\tilde{g}_{\varphi}\mathbf{V}_{\varphi}^{\dagger}\left( \begin{array}{cc}
	\Delta ^{\left( - \right)}&		i\Delta ^{\left( + \right)}\\
	-i\Delta ^{\left( + \right)}&		\Delta ^{\left( - \right)}\\
\end{array} \right) \mathbf{V}_{\varphi},~~\Delta ^{\left( \pm \right)}=\Delta \left( t_P,t'_{P'} \right) \pm \Delta \left( t'_{P'},t_P \right) .
\end{equation}

Thus, the action \eqref{3.4} written in terms of the fields \eqref{4.12} gets the form:      

\begin{equation}
\begin{split}
\label{4.22}
S & =  \frac{1}{2}\int\limits_0^{\pi}{d\varphi}\int\limits_P{dt_P}\bar{\bm{\uppsi}}_{\varphi}\left( i\partial _{t_P}+\varepsilon _{\varphi}\sigma ^z \right) \bm{\uppsi} _{\varphi} \cr & +\frac{i}{2}\int\limits_0^{\pi}{d\varphi}\int\limits_P{dt_P}\int\limits_{P'}{dt'_{P'}}\bar{\bm{\uppsi}}_{\varphi}\left( t_P \right) \mathbf{K}_{\varphi}\left( t_P,t'_{P'} \right) \bm{\uppsi} _{\varphi}\left( t'_{P'} \right)  \cr & +\frac{1}{2}\int\limits_0^{\pi}{d\varphi}\int\limits_P{dt_P\left( \mathbf{\bar{j}}_{\varphi}\bm{\uppsi} _{\varphi}+\bar{\bm{\uppsi}}_{\varphi}\mathbf{{j}}_{\varphi} \right)} .
\end{split}
\end{equation}

For the source term we followed the notation (see Eq. \eqref{3.6}): 

\begin{equation}
\label{4.23}
\mathbf{\bar{j}}_{\varphi}=\mathbf{f}_{\varphi}\mathbf{V}_{\varphi},~~~\mathbf{f}_{\varphi}=\left( f_{\varphi ,0},f_{\varphi ,1} \right) ,~~~f_{\varphi ,u}=\frac{1}{\sqrt{2\pi}}\sum_j{e^{i\varphi j}f_{j,u}}
\end{equation}

By minimizing Eq. \eqref{4.21} we get the classical equation:

\begin{equation}
\label{4.24}
\left( i\partial _{t_P}+\varepsilon _{\varphi}\sigma ^z \right) \bm{\uppsi}_{\varphi}^{cl.}\left( t_P \right) +i\int\limits_{P'}{dt'_{P'}}\mathbf{K}_{\varphi}\left( t_P,t'_{P'} \right) \bm{\uppsi}_{\varphi}^{cl.}\left( t'_{P'} \right) =-\mathbf{j}_{\varphi}\left( t_P \right) .
\end{equation}

The corresponding Green’s equation reads:

\begin{equation}
\label{4.25}
\left( i\partial _{t_P}+\varepsilon _{\varphi}\sigma ^z \right) \mathbf{L}_{\varphi}\left( t_P,t'_{P'} \right) +i\int\limits_{P''}{dt''_{P''}}\mathbf{K}_{\varphi}\left( t_P,t''_{P''} \right) \mathbf{L}_{\varphi}\left( t''_{P''},t'_{P'} \right) =\mathbf{I}\delta \left( t_P-t'_{P'} \right) .
\end{equation}

Having in hand the solution of the last equation and using Eq. \eqref{4.17} the generating functional gets the form

\begin{equation}
\label{4.26}
Z_{S}^{\left( 0 \right)}\left[ \mathbf{j} \right] =C_{E}^{-1}\exp \left[ -\frac{i}{2}\int\limits_0^{\pi}{d\varphi}\int\limits_{P^{\mathrm{'}}}{dt'_{P'}\int\limits_P{dt_P\mathbf{\bar{j}}_{\varphi}\left( t_P \right) \mathbf{L}_{\varphi}\left( t_P,t'_{P'} \right) \mathbf{j}\left( t'_{P'} \right)}} \right] .
\end{equation}

Taking into account \eqref{4.17} the generating functional can be rewritten in the form \eqref{3.11} with 

\begin{equation}
\label{4.27}
\Gamma _{jk,uv}\left( t_P,t'_{P'} \right) =\frac{1}{4\pi}\int\limits_0^{\pi}{d\varphi e^{i\varphi \left( j-k \right)}\left[ \mathbf{V}_{\varphi}\mathbf{L}_{\varphi}\left( t_P,t'_{P'} \right) \mathbf{V}_{\varphi}^{\dagger} \right] _{uv}}.
\end{equation}

The Green’s equation \eqref{4.25} with causal boundary conditions can be exactly solved. In Appenix A, we present the details of the calculation. Here it is enough to quote the result for $t_+=t'_+=t$:

\begin{equation}
\begin{split}
\label{4.28}
\mathbf{L}_{\varphi} & =-i\left[ \frac{1}{2}-\sin ^2\theta _{\varphi}\left( 1-e^{-\left| \tilde{g}_{\varphi} \right|\Gamma \left( t-t_{in.} \right)} \right) \right] \sigma ^z\\
&-\frac{i}{2}\frac{\left( \tilde{g}_{\varphi}\frac{\Gamma}{2} \right) ^2}{\varepsilon _{\varphi}^{2}+\left( \tilde{g}_{\varphi}\frac{\Gamma}{2} \right) ^2}\sin ^2\left( 2\theta _{\varphi} \right) \left( 1-e^{-\left| \tilde{g}_{\varphi} \right|\Gamma \left( t-t_{in.} \right)} \right) \left(\begin{array}{cc}
	\cos ^2\theta _{\varphi}&		0\\
	0&		\sin ^2\theta _{\varphi}\\
\end{array} \right) 
\end{split}
\end{equation}

By applying Eq. \eqref{3.14} it is straightforward to find the covariance matrix:              

\begin{equation}
\begin{split}
\label{4.29}
C_{jk,uv}\left( t \right) &=\frac{2}{\pi}\int\limits_0^{\pi}{d\varphi}\left( \begin{array}{cc}
	0&		\cos \left( \varphi \left( j-k \right) -2\theta _{\varphi} \right)\\
	-\cos \left( \varphi \left( j-k \right) +2\theta _{\varphi} \right)&		0\\
\end{array} \right) \\ &\times \left[ \frac{1}{2}-\sin ^2\theta _{\varphi}\left( 1-e^{-\frac{1}{2}\left| \tilde{g}_{\varphi} \right|\Gamma \left( t-t_{in.} \right)} \right) \right] \\
&  +\frac{1}{2\pi}\int\limits_0^{\pi}{d\varphi}\frac{\left( \tilde{g}_{\varphi}\frac{\Gamma}{2} \right) ^2}{\varepsilon _{\varphi}^{2}+\left( \tilde{g}_{\varphi}\frac{\Gamma}{2} \right) ^2}\sin ^2\left( 2\theta _{\varphi} \right) \left( 1-e^{-\left| \tilde{g}_{\varphi} \right|\Gamma \left( t-t_{in.} \right)} \right) \\
& \times \left( \begin{array}{cc}
	\sin \varphi \left( j-k \right)&		\cos \left( \varphi \left( j-k \right) -2\theta _{\varphi} \right) \cos \left( 2\theta _{\varphi} \right)\\
	-\cos \left( \varphi \left( j-k \right) +2\theta _{\varphi} \right) \cos \left( 2\theta _{\varphi} \right)&		\sin \varphi \left( j-k \right)\\
\end{array} \right) 
\end{split}
\end{equation}

At the limit $t\rightarrow \infty $, we get the covariance at the steady state:

\begin{equation}
\begin{split}
\label{4.30}
C_{jk,uv}^{\left( 0 \right)} & =\frac{1}{\pi}\int\limits_0^{\pi}{d\varphi}cos\left( 2\theta _{\varphi} \right) \left( \begin{array}{cc}
	0&		\cos \left( \varphi \left( j-k \right) -2\theta _{\varphi} \right)\\
	-\cos \left( \varphi \left( j-k \right) +2\theta _{\varphi} \right)&		0\\
\end{array} \right) \\
& +\frac{1}{2\pi}\int\limits_0^{\pi}{d\varphi}\frac{\left( \tilde{g}_{\varphi}\frac{\Gamma}{2} \right) ^2}{\varepsilon _{\varphi}^{2}+\left( \tilde{g}_{\varphi}\frac{\Gamma}{2} \right) ^2}\sin ^2\left( 2\theta _{\varphi} \right)  \\
&  \times \left( \begin{array}{cc}
	\sin \varphi \left( j-k \right)&		\cos \left( \varphi \left( j-k \right) -2\theta _{\varphi} \right) \cos \left( 2\theta _{\varphi} \right)\\
	-\cos \left( \varphi \left( j-k \right) +2\theta _{\varphi} \right) \cos \left( 2\theta _{\varphi} \right)&		\sin \varphi \left( j-k \right)\\
\end{array} \right) 
\end{split}
\end{equation}

The existence of the second term in the rhs of Eqs. \eqref{4.28} - \eqref{4.30} is due to the fact that the dissipation kernel in Eq. \eqref{4.21} has non-zero off-diagonal entries. An important observation is that both the functions \eqref{4.29} and \eqref{4.30} are non-analytic at the point $\left| h \right|=1$. Firstly, we neglect the off-diagonal contribution of the dissipation kernel by assuming that  $\left( g\Gamma /2 \right) \ll 1$. This assumption simplifies the results without qualitatively changing them. In this case the steady-state covariance reads:

\begin{equation}
\label{4.31}
C_{jk,uv}^{\left( 0 \right)}\simeq \frac{1}{\pi}\int\limits_0^{\pi}{d\varphi}\cos \left( 2\theta _{\varphi} \right) \left( \begin{array}{cc}
	0&		\cos \left( \varphi \left( j-k \right) -2\theta _{\varphi} \right)\\
	-\cos \left( \varphi \left( j-k \right) +2\theta _{\varphi} \right)&		0\\
\end{array} \right) 
\end{equation}

If the environment were absent ($\Gamma=0$) the corresponding result would have the form (see Eq. \ref{4.29}):

\begin{equation}
\label{4.32}
C_{jk,uv}^{\left( 0 \right)}\simeq \frac{1}{\pi}\int\limits_0^{\pi}{d\varphi}\left( \begin{array}{cc}
	0&		\cos \left( \varphi \left( j-k \right) -2\theta _{\varphi} \right)\\
	-\cos \left( \varphi \left( j-k \right) +2\theta _{\varphi} \right)&		0\\
\end{array} \right).
\end{equation}

This function represents the covariance at the ground state of the isolated system and it is non-analytic.  The integrals involved, are connected with the complete elliptic integral of the second kind that has a branch cut at $\left| h \right|=1$. This non-analytic behavior reflects the underlying quantum phase transition in the closed system. When the system is open the situation changes but non-analyticity persists. This is immediately confirmed by considering \eqref{4.31} for $j=k$:

\begin{equation}
\label{4.33}
C_{jj,uv}^{\left( 0 \right)}\simeq \frac{1}{\pi}\int\limits_0^{\pi}{d\varphi}\cos ^2\left( 2\theta _{\varphi} \right) \left( \begin{array}{cc}
	0&		1\\
	-1&		0\\
\end{array} \right) =\frac{1}{\pi}\int\limits_0^{\pi}{d\varphi}\frac{\left( h+\cos \varphi \right) ^2}{h^2+2h\cos \varphi +1}\left( \begin{array}{cc}
	0&		1\\
	-1&		0\\
\end{array} \right) 
\end{equation}

A simple calculation yields the result:
\begin{equation}
\label{4.34}
C_{jj,uv}^{\left( 0 \right)}\left( h \right) \simeq \left[ \frac{1}{2}\theta \left( 1-\left| h \right| \right) +\left( 1-\frac{1}{2h^2} \right) \theta \left( \left| h \right|-1 \right) \right] \left( \begin{array}{cc}
	0&		1\\
	-1&		0\\
\end{array} \right) 
\end{equation}
and

\begin{equation}
\label{4.35}
\frac{d}{d\left| h \right|}C_{jj,uv}^{\left( 0 \right)}\left( h \right) =\left( \begin{array}{cc}
	0&		1\\
	-1&		0\\
\end{array} \right) \times \left\{ \begin{array}{cc}
0,~~~~~~~~~~\left| h \right|<1\\
	{{1} /{\left| h \right|^3}},~~~\left| h \right|>1\\
\end{array} \right. 
\end{equation}

Thus, in the presence of dissipation the steady state’s covariance matrix possesses a finite discontinuity at the point $\left| h \right|=1$. This point of non-analyticity coincides with the critical point of the, well-known, quantum phase transition that characterizes the corresponding isolated system and its equivalent, the transverse Ising model.

This result indicates that, although differently, non-analyticity remains in the steady state of the open system despite the impact of the environment.

For $j-k=L\rightarrow +\infty $ a straightforward calculation yields the result:

\begin{equation}
\label{4.36}
C_{jk,uv}^{\left( 0 \right)}\left( h \right) \underset{L\rightarrow +\infty}{\simeq}\frac{1-h^2}{2h^2}\left( \begin{array}{cc}
	0&		h^L\theta \left( 1-\left| h \right| \right)\\
	\frac{1}{h^L}\theta \left( \left| h \right|-1 \right)&		0\\
\end{array} \right) 
\end{equation}
and
\begin{equation}
\label{4.37}
C_{jk,uv}^{\left( 0 \right)}\left( h \right) \underset{L\rightarrow -\infty}{\simeq}\frac{h^2-1}{2h^2}\left( \begin{array}{cc}
	0&		\frac{1}{h^{\left| L \right|}}\theta \left( \left| h \right|-1 \right)\\
	h^{\left| L \right|}\theta \left( 1-\left| h \right| \right)&		0\\
\end{array} \right) .
\end{equation}

Combining these expressions, we can infer that for $\left| L \right|\rightarrow \infty$ and $\left| \left| h \right|-1 \right|\rightarrow 0$ the steady state covariance behaves as \cite{Hoening2012,Eisert2010}:                                                                                                                                                  

\begin{equation}
\label{4.38}
C_{jk,uv}^{\left( 0 \right)}\left( h \right) \underset{L\rightarrow \infty ;\left| h \right|\rightarrow 1}{\thicksim}\left| \left| h \right|-1 \right|\exp \left( -\left| L \right|\left| \left| h \right|-1 \right| \right) .
\end{equation}

The last expression indicates the fact that the function $\frac{d}{d\left| h \right|}C_{jk,uv}^{\left( 0 \right)}\left( h \right) $ is characterized, at the limit $\left| h \right|\rightarrow 1$, by long range correlations the length of which has the form $\xi \sim \frac{1}{\left| \left| h \right|-1 \right|}\rightarrow \infty $.

As long as we are interested in functions of one time variable, as is the case of the steady state in the current Section, the results we obtained can also be derived through the Lindblad equation \cite{Horstmann2013}. In this direction, powerful techniques have been recently developed \cite{Zhang2022a,Zhang2022b} for solving quasi-free and quadratic Lindblad equations for bosonic and fermionic systems. The results of the present work, in which we have concentrated on the dissipation dynamics of a system with quadratic Hamiltonian, are in accordance with the results in Ref.  \cite{Zhang2022b} that refer to quasi-free Lindblad equations. In a forthcoming paper, we shall examine the dephasing dynamics of a quadratic system, a problem that lies in the framework of the quadratic Lindblad equation \cite{Zhang2022a}.

Our analysis has been conducted in the framework of the Markovian approximation, as encoded in Eq. \eqref{3.22}, which is the necessary condition for obtaining an analytical solution for Eq. \eqref{3.21} and the results presented in this Section. Non-Markovian dynamics, although not the subject of the current investigation, is of great interest both for theoretical and experimental reasons. Most of the time they are examined in the framework of master equations of the Lindblad type with time-dependent jump operators and decay rates \cite{Breuer2015}. In the framework of the reduced generating functional, the issue has been analyzed for a bosonic system through numerical techniques \cite{Kordas2018}. As expected from its structure, the behavior of the function \eqref{2.13} strongly depends on the spectral function of the environment. Consequently, one expects this dependence to be reflected in the analyticity properties of the stationary state \cite{Nagy2015}. As the characteristic feature of non-Markovianity is the exchange of information from the system to the environment and back, we don’t expect any change of the point of non-analyticity per se as this is connected with the zero of the energy gap \cite{Horstmann2013,Chen2011} at which the dynamics is almost Markovian \cite{Haikka2015,Wei2016}.  However, a change of the critical exponents is possible \cite{Nagy2015}.  Our results refer to the asymptotic limit $t \to \infty $ at which the main contribution to the function \eqref{2.13} comes from the region $t \sim t'$. Thus, for small enough coupling $g$, we expect the leading order behavior of our results to remain almost intact. 

\section{Conlusions}
\label{conclusions}
\setcounter{equation}{0}

We developed a framework for the construction of the time-dependent correlation functions of an open fermionic system for every number of time variables. It is a framework, that is based on the merge of the Feynman-Vernon’s technique with the Keldysh-Schwinger formalism on the ground of the coherent state path integrals. By integrating out the environmental degrees of freedom, we defined the generating functional through which every system’s correlation function can be derived by applying the appropriate functional derivatives. We presented the details of this construction in the case of a system that is quadratic when written in terms of Majorana variables. In the last Section of the work, we applied the proposed formalism to examine a model which is completely solvable when isolated namely, the transverse Ising model. We calculate the exact form of the covariance matrix at the steady state. We confirm the it is non-analytic at the same point at which the corresponding isolated model develops a quantum phase transition and we examine the specific form in which this non-analyticity is realized.

\section*{Appendix A}
\def\theequation{A.\arabic{equation}}
\setcounter{equation}{0}
\label{appendix:a}

In this Appendix we solve the Green’s equation \eqref{4.25} of the main text:

\begin{equation}
\label{a.1}
\left( i\partial _{t_P}+\varepsilon _{\varphi}\sigma ^z \right) \mathbf{L}_{\varphi}\left( t_P,t'_{P'} \right) +i\int\limits_{P''}{dt''_{P''}}\mathbf{K}_{\varphi}\left( t_P,t''_{P''} \right) \mathbf{L}_{\varphi}\left( t''_{P''},t'_{P'} \right) =\mathbf{I}\delta \left( t_P-t'_{P'} \right) .
\end{equation}

The solution we are interested for, must obey the boundary conditions:

\begin{equation}
\label{a.2}
\mathbf{L}\left( t_P=t_{in.+},t'_{P'} \right) =\mathbf{L}\left( t_P,t'_{P'}=t_{in.-} \right) =0,~~~\mathbf{L}\left( t_P,t'_{P'} \right) \underset{\left| t-t' \right|\rightarrow +\infty}{\rightarrow}0.
\end{equation}

The first demand expresses the causal character of propagation. The second one is connected with the decoherence encoded into the third term of Eq. \eqref{a.1}, the term that brings the influence of the environment into the propagation.

The dissipation kernel appearing in last equation has the form:                                                                                                        
\begin{equation}
\label{a.3}
{\tiny
 \mathbf{K}_{\varphi}=\tilde{g}_{\varphi}\left( \begin{array}{cc}
	\Delta \left( t_P,t'_{P'} \right) \cos ^2\theta _{\varphi}-\Delta \left( t'_{P'},t_P \right) \sin ^2\theta _{\varphi}&		-\frac{i}{2}\left( \Delta \left( t_P,t'_{P'} \right) +\Delta \left( t'_{P'},t_P \right) \right) \sin \left( 2\theta _{\varphi} \right)\\
	\frac{i}{2}\left( \Delta \left( t_P,t'_{P'} \right) +\Delta \left( t'_{P'},t_P \right) \right) \sin \left( 2\theta _{\varphi} \right)&		\Delta \left( t_P,t'_{P'} \right) \sin ^2\theta _{\varphi}-\Delta \left( t'_{P'},t_P \right) \cos ^2\theta _{\varphi}\\
\end{array} \right) .
}
\end{equation}

The solution of the Green’s equation is a matrix of Green’s functions: 

\begin{equation}
\label{a.4}
\mathbf{L}_{\varphi}=\left( \begin{array}{cc}
	L_{\varphi ,00}&		L_{\varphi ,01}\\
	L_{\varphi ,10}&		L_{\varphi ,11}\\
\end{array} \right) 
\end{equation}

The dissipation kernel splits the set of equations \eqref{a.1} to two groups of coupled equations:

\begin{equation}
\begin{split}
\label{a.5}
\partial _{t_P}L_{\varphi ,00}\left( t_P,t'_P \right) &-i\varepsilon _{\varphi}L_{\varphi ,00}\left( t_P,t'_P \right) +\int\limits_{P''}{{dt''}_P}K_{\varphi ,00}\left( t_P,{t''}_P \right) L_{\varphi ,00}\left( {t''}_P,t'_P \right)  \\
&+\int\limits_{P''}{{dt''}_P}K_{\varphi ,01}\left( t_P,{t''}_P \right) L_{\varphi ,10}\left( {t''}_P,t'_P \right) =-i\delta \left( t_p-t'_P \right)  \\
\partial _{t_P}L_{\varphi ,10}\left( t_P,t'_P \right) &+i\varepsilon _{\varphi}L_{\varphi ,10}\left( t_P,t'_P \right) +\int\limits_{P''}{{dt''}_P}K_{\varphi ,10}\left( t_P,{t^{''}}_P \right) L_{\varphi ,00}\left( {t''}_P,t'_P \right)  \\
&+\int\limits_{P''}{{dt''}_P}K_{\varphi ,11}\left( t_P,{t''}_P \right) L_{\varphi ,10}\left( {t''}_P,t'_P \right) =0
\end{split}
\end{equation}

and
\begin{equation}
\begin{split}
\label{a.6}
\partial _{t_P}L_{\varphi ,11}\left( t_P,t'_P \right) & +i\varepsilon _{\varphi}L_{\varphi ,11}\left( t_P,t'_P \right) +\int\limits_{P''}{{dt''}_P}K_{\varphi ,10}\left( t_P,{t''}_P \right) L_{\varphi ,01}\left( {t''}_P,t'_P \right) \\
&+\int\limits_{P''}{{dt''}_P}K_{\varphi ,11}\left( t_P,{t''}_P \right) L_{\varphi ,11}\left( {t''}_P,t'_P \right) =-i\delta \left( t_p-t'_P \right) \\
\partial _{t_P}L_{\varphi ,01}\left( t_P,t'_P \right) &-i\varepsilon _{\varphi}L_{\varphi ,01}\left( t_P,t'_P \right) +\int\limits_{P''}{{dt''}_P}K_{\varphi ,00}\left( t_P,{t''}_P \right) L_{\varphi ,01}\left( {t''}_P,t'_P \right) \\
&+\int\limits_{P''}{{dt''}_P}K_{\varphi ,01}\left( t_P,{t''}_P \right) L_{\varphi ,11}\left( {t''}_P,t'_P \right) =0.
\end{split}
\end{equation}

Restoring the explicit form of the Keldysh variables each of the groups \eqref{a.5} and \eqref{a.6} splits into two distinct groups of four coupled equations. As the calculations for each one of the $16$ groups have the same structure it is enough to present here only those groups that yield the result for $\mathbf{L}_{\varphi}\left( t_+,t'_+ \right) $.

In the following, and for reasons of simplicity, we have set $b=0$ and $\delta E=0$ in expressions \eqref{3.23} and we have used the abbreviations:

\begin{equation}
\label{a.7}
A_{\varphi}=\tilde{g}_{\varphi}\frac{\Gamma}{2}\cos \left( 2\theta _{\varphi} \right) ,~~~~B_{\varphi}=\tilde{g}_{\varphi}\frac{\Gamma}{2}\sin \left( 2\theta _{\varphi} \right).
\end{equation}

After these, the first tetrad of equations results from Eq. \eqref{a.5} and reads as follows:

\begin{equation}
\begin{split}
\label{a.8}
&\left( \partial _t-i\varepsilon _{\varphi}+A_{\varphi} \right) L_{\varphi ,00}\left( t_+,t_+ \right) +\Gamma \tilde{g}_{\varphi}\sin ^2\theta _{\varphi}L_{\varphi ,00}\left( t_-,t_+ \right) \\ &~~~~ -iB_{\varphi}\left( L_{\varphi ,10}\left( t_+,t_+ \right) -L_{\varphi ,10}\left( t_-,t_+ \right) \right) =-i\delta \left( t-t' \right),  \\
&\Gamma \tilde{g}_{\varphi}\cos ^2\theta _{\varphi}L_{\varphi ,00}\left( t_+,t_+ \right) +\left( \partial _t-i\varepsilon _{\varphi}-A_{\varphi} \right) L_{\varphi ,00}\left( t_-,t_+ \right) \\ &~~~~  -iB_{\varphi}\left( L_{\varphi ,10}\left( t_+,t_+ \right) -L_{\varphi ,10}\left( t_-,t_+ \right) \right) =0, \\
&iB_{\varphi}\left( L_{\varphi ,00}\left( t_+,t_+ \right) -L_{\varphi ,00}\left( t_-,t_+ \right) \right) +\left( \partial _t+i\varepsilon _{\varphi}-A_{\varphi} \right) L_{\varphi ,10}\left( t_+,t_+ \right) \\ &~~~~ +\Gamma \tilde{g}_{\varphi}\cos ^2\theta _{\varphi}L_{\varphi ,10}\left( t_-,t_+ \right) =0,\\
&iB_{\varphi}\left( L_{\varphi ,00}\left( t_+,t_+ \right) -L_{\varphi ,00}\left( t_-,t_+ \right) \right) +\Gamma \tilde{g}_{\varphi}\sin ^2\theta _{\varphi}L_{\varphi ,10}\left( t_+,t_+ \right) \\ &~~~~ +\left( \partial _t+i\varepsilon _{\varphi}+A_{\varphi} \right) L_{\varphi ,10}\left( t_-,t_+ \right) =0.
\end{split}
\end{equation}

The set of Eqs. \eqref{a.8} can be exactly solved. Without any boundary restriction it is a straightforward exercise to find:

\begin{equation}
\begin{split}
\label{a.9}
L_{\varphi ,00}\left( t_+,t'_+ \right) &=-i\cos ^2\theta _{\varphi}e^{i\varepsilon _{\varphi}\left( t-t' \right) -\frac{1}{2}\left| \tilde{g}_{\varphi} \right|\Gamma \left( t-t' \right)}\theta \left( t-t' \right) \\&+i\sin ^2\theta _{\varphi}e^{i\varepsilon _{\varphi}\left( t-t' \right) -\frac{1}{2}\left| \tilde{g}_{\varphi} \right|\Gamma \left( t'-t \right)}\theta \left( t'-t \right)  \\
&-\frac{i}{2}\frac{\left( \tilde{g}_{\varphi}\frac{\Gamma}{2} \right) ^2}{\varepsilon _{\varphi}^{2}+\left( \tilde{g}_{\varphi}\frac{\Gamma}{2} \right) ^2}\sin ^2\left( 2\theta _{\varphi} \right) \cos ^2\theta _{\varphi}e^{-\frac{1}{2}\left| \tilde{g}_{\varphi} \right|\Gamma \left| t-t' \right|} \\ & \times \left( \cos \left( \varepsilon _{\varphi}\left| t-t' \right| \right) +\tilde{g}_{\varphi}\frac{\Gamma}{2}\frac{\sin \left( \varepsilon _{\varphi}\left| t-t' \right| \right)}{\varepsilon _{\varphi}} \right) 
\end{split}
\end{equation}
and
\begin{equation}
\label{a.10}
L_{\varphi ,10}\left( t_+,t'_+ \right) =-\frac{1}{2}e^{-\frac{1}{2}\left| \tilde{g}_{\varphi} \right|\Gamma \left( t-t' \right)}\tilde{g}_{\varphi}\frac{\Gamma}{2}\frac{\sin \varepsilon _{\varphi}\left( t-t' \right)}{\varepsilon _{\varphi}}\theta \left( t-t' \right) .
\end{equation}

The second group we are interested for is connected with Eq. \eqref{a.6} and assumes the form:

\begin{equation}
\begin{split}
\label{a.11}
&\left( \partial _t+i\varepsilon _{\varphi}-A_{\varphi} \right) L_{\varphi ,11}\left( t_+,t'_+ \right) +\Gamma \tilde{g}_{\varphi}\cos ^2\theta _{\varphi}L_{\varphi ,11}\left( t_-,t'_+ \right)  \\ &~~~~ +iB_{\varphi}\left( L_{\varphi ,01}\left( t_+,t'_+ \right) -L_{\varphi ,01}\left( t_-,t'_+ \right) \right) =-i\delta \left( t-t' \right)  ,  \\
&\Gamma \tilde{g}_{\varphi}\sin ^2\theta _{\varphi}L_{\varphi ,11}\left( t_+,t'_+ \right) +\left( \partial _t+i\varepsilon _{\varphi}+A_{\varphi} \right) L_{\varphi ,11}\left( t_-,t'_+ \right)  \\ &~~~~ +iB_{\varphi}\left( L_{\varphi ,01}\left( t_+,t'_+ \right) -L_{\varphi ,01}\left( t_-,t'_+ \right) \right) =0, \\
&-iB_{\varphi}\left( L_{\varphi ,11}\left( t_+,t'_+ \right) -L_{\varphi ,11}\left( t_-,t'_+ \right) \right) +\left( \partial _t-i\varepsilon _{\varphi}+A_{\varphi} \right) L_{\varphi ,10}\left( t_+,t'_+ \right) \\ &~~~~+\Gamma \tilde{g}_{\varphi}\sin ^2\theta _{\varphi}L_{\varphi ,01}\left( t_-,t'_+ \right) =0,\\
&-iB_{\varphi}\left( L_{\varphi ,11}\left( t_+,t'_+ \right) -L_{\varphi ,11}\left( t_-,t'_+ \right) \right)  +\Gamma \tilde{g}_{\varphi}\cos ^2\theta _{\varphi}L_{\varphi ,01}\left( t_+,t'_+ \right) \\ &~~~~ +\left( \partial _t-i\varepsilon _{\varphi}-A_{\varphi} \right) L_{\varphi ,01}\left( t_-,t_+ \right) =0.
\end{split}
\end{equation}

The solution of this system can be deduced from the solution of the system \eqref{a.8} by observing that the coefficients in \eqref{a.11} can be produced from the coefficients in (A.10) by making the replacement $\theta _{\varphi}\rightarrow \theta _{\varphi}\pm {{\pi} /{2}}$. This makes easy to find:

\begin{equation}
\begin{split}
\label{a.12}
L_{\varphi ,11}\left( t_+,t'_+ \right) & =-i\sin ^2\theta _{\varphi}e^{i\varepsilon _{\varphi}\left( t-t' \right) -\frac{1}{2}\left| \tilde{g}_{\varphi} \right|\Gamma \left( t-t' \right)}\theta \left( t-t' \right) \\ & +i\cos ^2\theta _{\varphi}e^{i\varepsilon _{\varphi}\left( t-t' \right) -\frac{1}{2}\left| \tilde{g}_{\varphi} \right|\Gamma \left( t'-t \right)}\theta \left( t'-t \right)  \\ & -\frac{i}{2}\frac{\left( \tilde{g}_{\varphi}\frac{\Gamma}{2} \right) ^2}{\varepsilon _{\varphi}^{2}+\left( \tilde{g}_{\varphi}\frac{\Gamma}{2} \right) ^2}\sin ^2\left( 2\theta _{\varphi} \right) \sin ^2\theta _{\varphi}e^{-\frac{1}{2}\left| \tilde{g}_{\varphi} \right|\Gamma \left| t-t' \right|}\\ & \times \left( \cos \left( \varepsilon _{\varphi}\left| t-t' \right| \right) +\tilde{g}_{\varphi}\frac{\Gamma}{2}\frac{\sin \left( \varepsilon _{\varphi}\left| t-t' \right| \right)}{\varepsilon _{\varphi}} \right) 
\end{split}
\end{equation}
and

\begin{equation}
\label{a.13}
L_{\varphi ,01}\left( t_+,t'_+ \right) =L_{\varphi ,10}\left( t_+,t'_+ \right).
\end{equation}

As discussed in the main text (see Eqs.  \eqref{4.23} and \eqref{4.24}) the Green’s matrix $\mathbf{L}$ propagates a two-component fermion field:

\begin{equation}
\label{a.14}
\psi _{\varphi ,u}^{cl.}\left( t_P \right) =\sum_{\nu}{\int\limits_{P'}{dt'_{P'}}}L_{\varphi ,u\nu}\left( t_P,t'_{P'} \right) j_{\varphi ,\nu}\left( t'_{P'} \right) .
\end{equation}

One of the components is propagating forward in time, while the second one, backwards.
Based on this observation, we construct, the appropriate for the present problem, Green’s functions by imposing the following boundary conditions:

\begin{equation}
\begin{split}
\label{a.15}
& L_{\varphi ,00}\left( t_+=t_{in.+},t'_+ \right) =0,~L_{\varphi ,01}\left( t_+=t_{in.+},t'_+ \right) =0 \\ & L_{\varphi ,10}\left( t_+,t'_+=t_{in.+} \right) =0,~L_{\varphi ,11}\left( t_+,t'_+=t_{in.+} \right) =0.
\end{split}
\end{equation}

To obtain these functions we can add at the Green’s functions we found, the general solution of the corresponding homogenous equation

\begin{equation}
\label{a.16}
L_{\varphi ,u\nu}\rightarrow L_{\varphi ,u\nu}+\sum_{j=1}^4{a_{\varphi ,u\nu}^{\left( j \right)}}e^{ip_jt},~~~p_j=\pm \varepsilon _{\varphi}\pm i\frac{1}{2}\left| \tilde{g}_{\varphi} \right|\Gamma 
\end{equation}
and determine the coefficients to satisfy \eqref{a.15} together with $L_{\varphi ,u\nu}\underset{\left| t-t' \right|\rightarrow +\infty}{\rightarrow}0$. In this way we find:

\begin{equation}
\begin{split}
\label{a.17}
L_{\varphi ,00}\left( t_+,t'_+ \right) & =-ie^{i\varepsilon _{\varphi}\left( t-t' \right) -\frac{1}{2}\left| \tilde{g}_{\varphi} \right|\Gamma \left| t-t' \right|}\theta \left( t-t' \right) \\ & +iF_{\varphi ,00}\left( t-t' \right) \left[ e^{-\frac{1}{2}\left| \tilde{g}_{\varphi} \right|\Gamma \left| t-t' \right|}-e^{-\frac{1}{2}\left| \tilde{g}_{\varphi} \right|\Gamma \left( t+t'-2t_{in.} \right)} \right] 
\end{split}
\end{equation}
and
\begin{equation}
\begin{split}
\label{a.18}
L_{\varphi ,11}\left( t_+,t'_+ \right) &=ie^{i\varepsilon _{\varphi}\left( t-t' \right) -\frac{1}{2}\left| \tilde{g}_{\varphi} \right|\Gamma \left| t-t' \right|}\theta \left( t'-t \right)\\&-iF_{\varphi ,11}\left( t-t' \right) \left[ e^{-\frac{1}{2}\left| \tilde{g}_{\varphi} \right|\Gamma \left| t-t' \right|}-e^{-\frac{1}{2}\left| \tilde{g}_{\varphi} \right|\Gamma \left( t+t'-2t_{in.} \right)} \right].
\end{split}
\end{equation}

The abbreviations in these expressions read as follows:

\begin{equation}
\begin{split}
\label{a.19}
F_{\varphi ,00}&=\sin ^2\theta _{\varphi}e^{i\varepsilon _{\varphi}\left( t-t' \right)}
\\&-\frac{1}{2}\frac{\left( \tilde{g}_{\varphi}\frac{\Gamma}{2} \right) ^2}{\varepsilon _{\varphi}^{2}+\left( \tilde{g}_{\varphi}\frac{\Gamma}{2} \right) ^2}\sin ^2\left( 2\theta _{\varphi} \right) \cos ^2\theta _{\varphi} \left( \cos \left( \varepsilon _{\varphi}\left| t-t' \right| \right) +\tilde{g}_{\varphi}\frac{\Gamma}{2}\frac{\sin \left( \varepsilon _{\varphi}\left| t-t' \right| \right)}{\varepsilon _{\varphi}} \right) 
\end{split}
\end{equation}
and
\begin{equation}
\begin{split}
\label{a.20}
F_{\varphi ,11}&=\sin ^2\theta _{\varphi}e^{i\varepsilon _{\varphi}\left( t-t' \right)}
\\&+\frac{1}{2}\frac{\left( \tilde{g}_{\varphi}\frac{\Gamma}{2} \right) ^2}{\varepsilon _{\varphi}^{2}+\left( \tilde{g}_{\varphi}\frac{\Gamma}{2} \right) ^2}\sin ^2\left( 2\theta _{\varphi} \right) \sin ^2\theta _{\varphi} \left( \cos \left( \varepsilon _{\varphi}\left| t-t' \right| \right) +\tilde{g}_{\varphi}\frac{\Gamma}{2}\frac{\sin \left( \varepsilon _{\varphi}\left| t-t' \right| \right)}{\varepsilon _{\varphi}} \right) .
\end{split}
\end{equation}

In the same way we find

\begin{equation}
\begin{split}
\label{a.21}
L_{\varphi ,01}\left( t_+,t'_+ \right)  =-\frac{1}{2}e^{-\frac{1}{2}\left| \tilde{g}_{\varphi} \right|\Gamma \left( t-t' \right)}\tilde{g}_{\varphi}\frac{\Gamma}{2}\frac{\sin \varepsilon _{\varphi}\left( t-t' \right)}{\varepsilon _{\varphi}}\theta \left( t-t' \right) 
\end{split}
\end{equation}
and
\begin{equation}
\begin{split}
\label{a.21b}
L_{\varphi ,10}\left( t_+,t'_+ \right) & =\frac{1}{2}e^{-\frac{1}{2}\left| \tilde{g}_{\varphi} \right|\Gamma \left| t-t' \right|}\tilde{g}_{\varphi}\frac{\Gamma}{2}\frac{\sin \varepsilon _{\varphi}\left| t-t' \right|}{\varepsilon _{\varphi}}\theta \left( t'-t \right)  \\ & -e^{-\frac{1}{2}\left| \tilde{g}_{\varphi} \right|\Gamma \left| t-t' \right|}\tilde{g}_{\varphi}\frac{\Gamma}{2}\frac{\sin \varepsilon _{\varphi}\left| t-t' \right|}{\varepsilon _{\varphi}}\left[ e^{-\frac{1}{2}\left| \tilde{g}_{\varphi} \right|\Gamma \left| t-t' \right|}-e^{-\frac{1}{2}\left| \tilde{g}_{\varphi} \right|\Gamma \left( t+t'-2t_{in.} \right)} \right] .
\end{split}
\end{equation}

Combining these results, we find that for $t_+=t'_+=t$:

\begin{equation}
\begin{split}
\label{a.23}
\mathbf{L}_{\varphi} & =-i\left[ \frac{1}{2}-\sin ^2\theta _{\varphi}\left( 1-e^{-\left| \tilde{g}_{\varphi} \right|\Gamma \left( t-t_{in.} \right)} \right) \right] \sigma ^z 
\\ & -\frac{i}{2}\frac{\left( \tilde{g}_{\varphi}\frac{\Gamma}{2} \right) ^2}{\varepsilon _{\varphi}^{2}+\left( \tilde{g}_{\varphi}\frac{\Gamma}{2} \right) ^2}\sin ^2\left( 2\theta _{\varphi} \right) \left( 1-e^{-\left| \tilde{g}_{\varphi} \right|\Gamma \left( t-t_{in.} \right)} \right) \left(\begin{array}{cc}
	\cos ^2\theta _{\varphi}&		0\\
	0&		\sin ^2\theta _{\varphi}\\
\end{array} \right) 
\end{split}
\end{equation}

%\section*{Acknowledgments}

\endpaper
%\end{document}

\end{document}